\documentclass[aps, prl, 10pt, twocolumn, groupedaddress, superscriptaddress, nofootinbib]{revtex4-2}

\usepackage{graphicx}
\usepackage[british]{babel} % document
\usepackage{amsmath,amssymb,amsfonts}
\usepackage[colorlinks,citecolor=blue,linkcolor=blue,urlcolor=blue, breaklinks=true]{hyperref}
\usepackage{color}
\usepackage[normalem]{ulem}
\usepackage{float}
\usepackage{simpler-wick}
\begin{document}

\title{Additive Manufacturing for Advanced Quantum Technologies}
%\author{Feiran Wang}
\author{F. Wang}
\thanks{The first two authors contributed equally to this work}
\affiliation{Centre for Additive Manufacturing, Faculty of Engineering, University of Nottingham, University Park, Nottingham, NG7 2RD, UK}
%\author{Nathan Cooper}
\author{N. Cooper}
\thanks{The first two authors contributed equally to this work}
\affiliation{School of Physics and Astronomy, University of Nottingham, University Park, Nottingham, NG7 2RD, UK}
%\author{Lucia Hackerm\"{u}ller}
%\author{David Johnson}
\author{D. Johnson}
\affiliation{School of Physics and Astronomy, University of Nottingham, University Park, Nottingham, NG7 2RD, UK}
\author{B. Hopton}
\affiliation{School of Physics and Astronomy, University of Nottingham, University Park, Nottingham, NG7 2RD, UK}
%\author{T Mark Fromhold}
\author{T. M. Fromhold}
\affiliation{School of Physics and Astronomy, University of Nottingham, University Park, Nottingham, NG7 2RD, UK}
\author{R. Hague}
\affiliation{Centre for Additive Manufacturing, Faculty of Engineering, University of Nottingham, University Park, Nottingham, NG7 2RD, UK}
%\author{Ashleigh Murray}
\author{A. Murray}
\affiliation{Centre for Additive Manufacturing, Faculty of Engineering, University of Nottingham, University Park, Nottingham, NG7 2RD, UK}
%\author{Ryan McMullen}
\author{R. McMullen}
\affiliation{Centre for Additive Manufacturing, Faculty of Engineering, University of Nottingham, University Park, Nottingham, NG7 2RD, UK}
%\author{Richard Hague}
%\author{Lyudmila Turyanska}
\author{L. Turyanska}
\affiliation{Centre for Additive Manufacturing, Faculty of Engineering, University of Nottingham, University Park, Nottingham, NG7 2RD, UK}
\author{L. Hackerm\"{u}ller}
\thanks{Corresponding author: \\ lucia.hackermuller@nottingham.ac.uk\\}
\affiliation{School of Physics and Astronomy, University of Nottingham, University Park, Nottingham, NG7 2RD, UK}
\date{March 2025}

\begin{abstract} \noindent The development of quantum technology has opened up exciting opportunities to revolutionize computing and communication, timing and navigation systems, enable non-invasive imaging of the human body, and probe fundamental physics with unprecedented precision. Alongside these advancements has come an increase in experimental complexity and a correspondingly greater dependence on compact, efficient and reliable hardware. The drive to move quantum technologies from laboratory prototypes to portable, real-world instruments has incentivized miniaturization of experimental systems relating to a strong demand for smaller, more robust and less power-hungry quantum hardware and for increasingly specialized and intricate components. Additive manufacturing, already heralded as game-changing for many manufacturing sectors, is especially well-suited to this task owing to the comparatively large amount of design freedom it enables and its ability to produce intricate three-dimensional forms and specialized components. Herein we review work conducted to date on the application of additive manufacturing to quantum technologies, discuss the current state of the art in additive manufacturing in optics, optomechanics, magnetic components and vacuum equipment, and consider pathways for future advancement. We also give an overview of the research and application areas most likely to be impacted by the deployment of additive manufacturing techniques within the quantum technology sector.
\end{abstract}

\maketitle

\section{Introduction}

Quantum technologies (QT) are expected to have a transformative impact on both research and society, enabling breakthroughs in the study of both fundamental and emergent phenomena as well as underpinning technologies with immediate societal benefits. Some QT devices, such as quantum computers and quantum simulators\,\cite{qcomp1,qcomp2,universal_qsim}, which are expected to drastically expand our ability to understand and simulate a range of complex processes across the natural sciences\,\cite{qsim1,qsim2,qsim3,qsim4}, can perform these functions as static devices. Other quantum technologies, such as those associated with quantum cryptography\,\cite{qcrypt1,qcrypt2}, could increase their impact further through portability but are also valuable as fixed installations. Manufacturing requirements for these technologies are currently centered around scalability, stability and rapid prototyping.

However, one of the aspects of QT that is closest to realizing major societal and commercial benefits is quantum sensing. Devices such as atomic clocks\,\cite{bize2005cold,optclocks,Grotti2024,Loh2025}, cold atom gravimeters\,\cite{biggrav1,biggrav2,Ruan2024,Cassens2025, Elliot2023}, rotation sensors\,\cite{atomic_gyro1,atomic_gyro2,Meng2024,Salducci2024} and quantum magnetometers\,\cite{MIT,fabricant2023,coldmag} are presenting opportunities for us to see underground\,\cite{submap,Shettell2024,Diament2024}, navigate without external signals\,\cite{inert_nav1,inert_nav2} and non-invasively image the structure and function of the human body in real time\,\cite{SPMIC,OPMneuroimage,Guo_2023,Zhao2023,dfmagnetometer}. To be of use, these sensing technologies need to be deployed outside the laboratory, in proximity to the subjects they are supposed to be sensing. In recent years this has created a strong drive towards miniaturization, with a clear need for compact, portable versions of many quantum sensors\,\cite{gravimeter1,gravimeter2,PortableCAG,portmag1,minigyro}.  This need is strengthened further by the recent surge in interest in deploying QT hardware in space\,\cite{QPinSpace}; whether for Earth-monitoring applications\,\cite{Earth_obs1,Earth_obs2}, precisions tests and astronomical observations\,\cite{equivtest2,spaceint,planet_obs} or searches for new physics\,\cite{AION}, drastic reductions in size, weight and power consumption (SWAP) are essential for space-based QT. Miniaturization, integration and mass-reduction are therefore key aspects of any manufacturing strategy for quantum sensing technologies.

%This drive for miniaturisation provides perhaps the largest and most obvious incentive for the deployment of additive manufacturing (AM) within QT. With laboratory prototypes of QT devices often filling an entire room and requiring kW of power, there is a need for drastic reductions in the size, weight and power consumption (SWAP), as well as the cost, of QT hardware. 

%The most obvious incentive for the deployment of additive manufacturing (AM) within QT is the opportunity to drastically reduce the cost, as well as the size, weight and power consumption (SWAP), of QT hardware. With laboratory prototypes of devices such as atomic clocks \cite{bize2005cold}, cold atom gravimeters \cite{biggrav1,biggrav2}, rotation sensors \cite{atomic_gyro1,atomic_gyro2} and the like often filling an entire room and requiring kW of power, the need for SWAP reductions to enable deployment outside the lab is clear. This need is strengthened further by the recent surge in interest in deploying QT hardware in space \cite{QPinSpace}; whether for Earth-monitoring applications \cite{Earth_obs1,Earth_obs2}, precisions tests and astronomical observations \cite{equivtest2,spaceint,planet_obs} or searches for new physics \cite{AION}, drastic reductions in SWAP are essential for space-based QT. 
\begin{figure}[h!]
    \centering
    \includegraphics[width=0.5\textwidth]{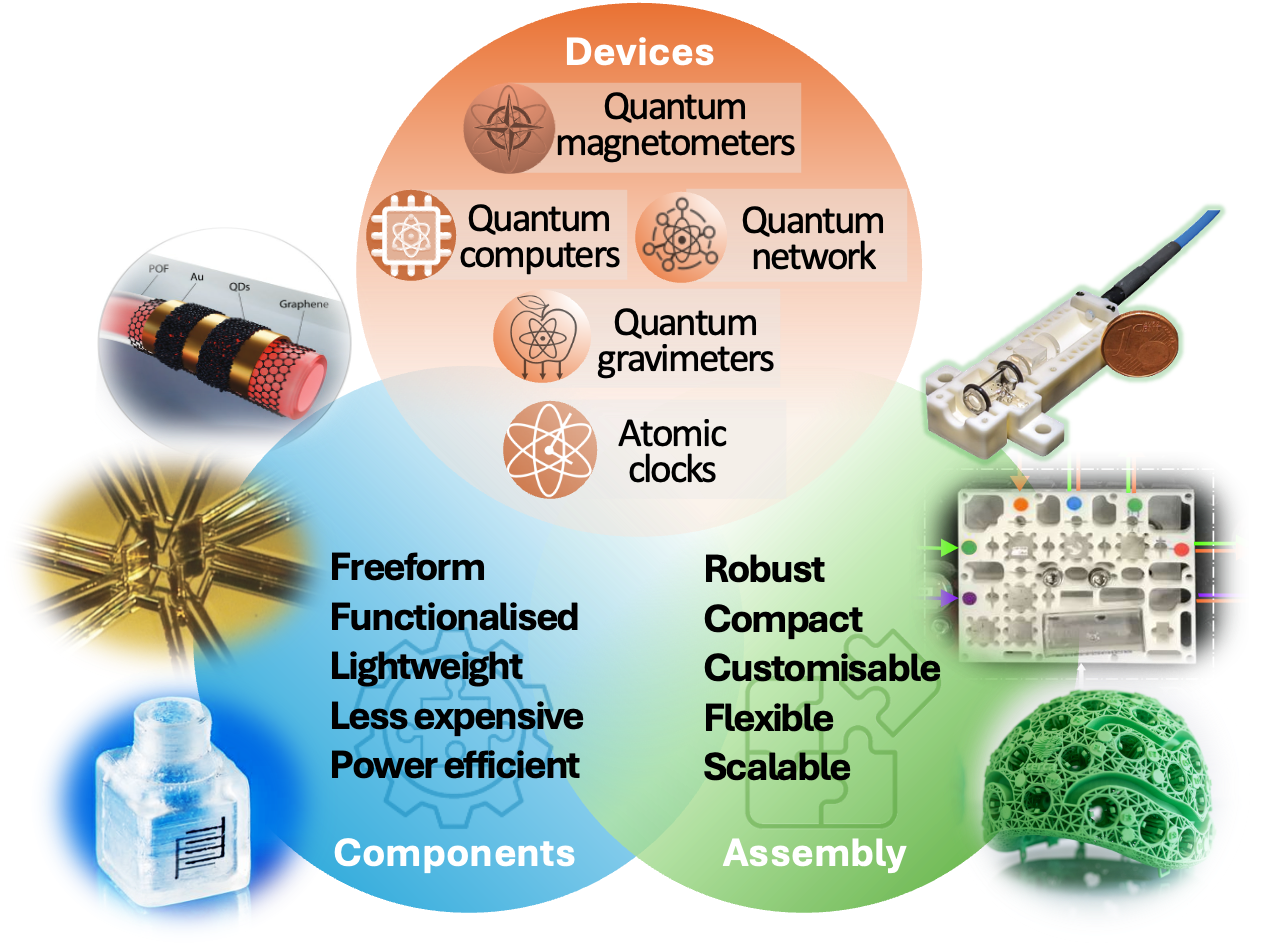}
    \caption{Examples of quantum technology devices and the benefits of AM for component manufacturing and device assembly. The insets are (clockwise from bottom left): a functionalized additively manufactured (AM) vapour cell\,\cite{wang2024additive}, a printed ion trap for quantum computing\,\cite{xu20233d}, an AM photodetector printed directly onto an optical fibre tip\,\cite{kara2023conformal}, a spectroscopic reference system housed in an AM ceramic mount\,\cite{ceramref}, optical apparatus for atom trapping and cooling integrated into an AM polymer framework\,\cite{optamot} and an AM helmet for positioning of quantum sensors in a wearable magnetoencephalography system\,\cite{Hill_2020}.
    }
    \label{fig:summary}
\end{figure}

%Another motivation is to improve the robustness and stability of QT experiments. Laboratory devices are carefully shielded from destabilising enironmental effects, such as vibrations and fluctuations of temperature or humidity, are not subject to significant physical strain or extreme conditions that might degrade or destroy hardware, and are operated only by expert users with a high level of technical understanding. Deployment of QT in real-world applications requires highly-stable, sturdy hardware that can withstand vibrations, accelerations, temperature changes and gradients and that can be operated successfully by end users who may have little or no knowledge of the underlying workings of the devices they employ. The flexibility of AM processes offers opportunities to adapt hardware to meet these criteria while still preserving its core function. 

The last decade has seen dramatic growth of quantum technology sectors and the importance of quantum technologies for global economies -- with additive manufacturing (AM) set to enable an impressive improvement in precision and robustness. There are several promising examples of additively manufactured deployable components, with lighter weight, tailored designs allowing fabrication of small numbers of intricate, custom components and embedded functionalities; these are enabled by combining different AM technologies to manufacture a specific part \cite{optamot} and integrating functional devices with structural architectures and active parts \cite{AM_VC}. Here we review recent advancements in manufacturing of QTs enabled by AM, and discuss future perspectives and directions the field is likely to take. 

%figure* does float -- not ideal, but fixes for now

\begin{figure*}%[t]
    \centering
    \includegraphics[width=0.9\textwidth]{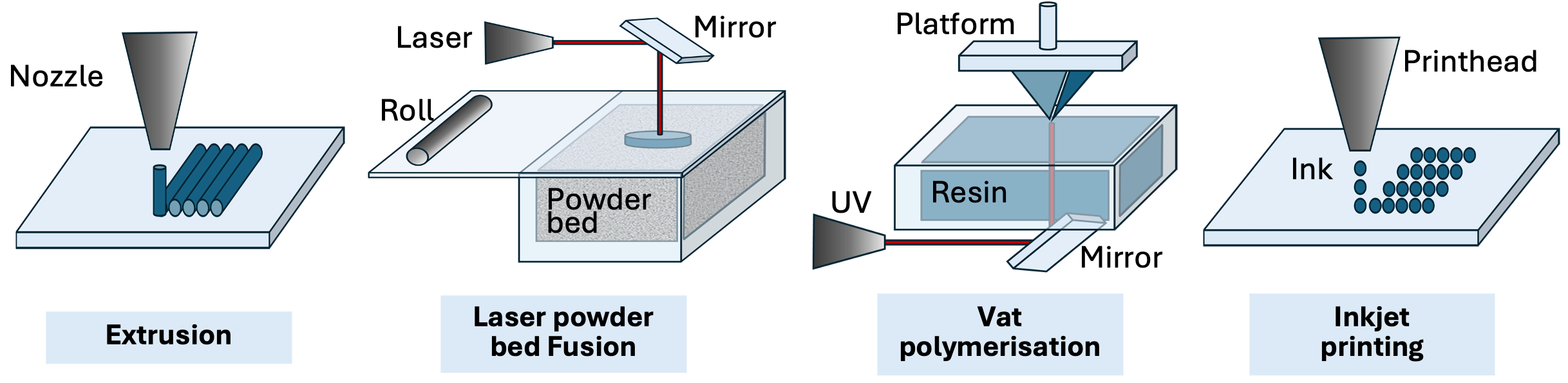}
    \caption{Schematics of different additive manufacturing technologies, which have been applied for quantum technologies.}
    \label{fig:am_methods}
\end{figure*}

Figure\,\ref{fig:summary} summarizes the benefits AM can offer for various QT devices, both at the level of component fabrication and system assembly. Assembly can be facilitated through the use of AM structural elements, whose main requirements are of shape and physical integrity. AM components include optics, which require accurate shaping, low surface roughness and appropriate transparency and refractive properties, as well as active components that integrate conductive pathways and functional electronics within AM materials. Below we give an overview of each of these areas, as well as an assessment of likely future directions and the demonstrated and expected impacts of AM within QT. First, however, we give an overview of the different AM methods used in their production and a brief discussion of design methodology for AM.

%The QT applications of AM to date can broadly be divided into three categories: structural components, whose main requirements are of shape and physical integrity; optics, which require accurate shaping, low surface roughness and appropriate transparency and refractive properties; and active components that integrate conductive pathways and functional electronics within AM components. Below we give an overview of each of these areas, as well as a summary of the likely future directions and impacts of AM within QT. First, however, we give an overview of the different AM methods used in their production and a brief discussion of design methodology for AM. 

%Motivation: portable quantum technologies, space applications

%prospects for miniaturisation, low SWAP

%optimisation (Chris magnetic fields, vacuum chambers vs. external pressure)

\section{Additive Manufacturing methods}

%Initially established as a rapid prototyping tool in the 1980s, AM has evolved into several distinct technologies, each with their own material compatibilities, geometrical capabilities, and mechanical properties to consider. Depending on the printing technique used, printed materials can encompass polymers, metals, ceramics, and composites, and print scale can vary from sub-micron (two-photon polymerisation) to tens of meters (material extrusion of concrete [2]). The recent development of AM compatible functional materials has further improved and expanded printing capability to include electronic and optically active devices. These functionalities, combined with the unique geometric freedoms, customisability, and scalability of AM offer a distinct advantage in the manufacturing of quantum devices.
Several AM techniques have been applied in the production and improvement of quantum devices, as summarized in Figure\,\ref{fig:am_methods}. Extrusion based methods such as fused deposition modeling (FDM) have been  used typically in prototyping and manufacturing small structural parts\,\cite{Staerkind_2023}. FDM involves the heating of a thermoplastic filament which is extruded through a nozzle onto a build plate. Similarly, direct writing extrudes high viscosity pastes, polymers and ceramic slurries which are then either dried or cured photochemically. While benefiting from ease of use, cost-effectiveness, and speed, extruded pieces can be prone to structural weakness between printed layers and have a resolution that is limited to nozzle dimensions ($>50 \,\mu$m). 

Laser powder bed fusion (LPBF) utilizes a laser or electron beam to fuse together layers of powdered metal, ceramic, or polymeric materials. The print bed is then lowered and fresh powder spread using a recoating lade or roller for the next layer to be fused. The resolution of this technique is controlled by both the laser spot size and the size of the powder particles. The roughness of the surface is able to reach 10\,$\mu$m\ after post-process polishing\,\cite{sigel2017miniaturization}. The benefits of this technique include the ability to produce miniaturized, lightweight, complex geometries, ideally suited to improving device portability and power density. One drawback of this approach is that substantial heat is dissipated in the build material, which can restrict printing speed and/or cause damaging thermal stresses within the part under construction; designs intended for fabrication via LPBF must account for this to ensure adequate build quality. LPBF has been used to produce a Ni-5Mo-15Fe magnetic shield, a Ti-6Al-4V vacuum flange\,\cite{birmingham_flange}, and an AlSi10Mg coil for magneto-optical atom trapping\,\cite{saint_coils}.

Vat polymerization techniques utilize light sources to photochemically cure polymeric resins within a reservoir. Stereolithogray (SLA), digital light processing (DLP), masked stereolithography (mSLA), micro-stereolithography, and emerging technologies such as volumetric additive manufacturing (VAM)\,\cite{shusteff2017one}, and two-photon polymerization (2PP) all fall into this category. These techniques have the advantage of rapidly producing low cost and high-resolution prints, enabling simulated results to be directly implemented into test mounts and expediting design optimization. 
For example, SLA was employed to rapidly prototype and optimize a monolithic polymer mount for the optical systems required in atom trapping experiments; by directly implementing the functional requirements as efficiently as possible, this device offered significant miniaturization and enhanced stability \cite{optamot}.  Two-photon polymerization processes have improved spatial selectivity, enabling sub-micron resolution, and are well suited for miniaturized, highly-resolved prints \cite{2pp1,2pp2}. 
Ceramics and nanoparticles have been added to resins to incorporate additional functionalities to printed pieces, such as conductivity or enhanced mechanical strength \cite{im2023strategies}.  One such resin has been formulated to include 50 wt\% fumed silica nanoparticles, and was used to manufacture dense, amorphous glass structures with high optical quality\,\cite{kotz2017three}. This material has been used to produce high quality micro-optics and vapor cells capable of ultra-high vacuum levels\,\cite{wang2024additive}, allowing for Doppler-free spectroscopy and laser frequency stabilization. 

Inkjet printing, also known as material jetting, is a well-established technique in conventional graphic printers. Various droplet generation methods are used to produce a stream of, or single, ink droplets that are expelled towards a substrate, allowing for selective patterning without the need for masks. Inkjet printing is capable of printing a wide range of functional materials, with resolution controlled by droplet size and able to reach less than 50 $\mu$m\,\cite{nelson2024off}. Recent research work aimed at expanding the range of compatible inks to include optical, electronic and optoelectronically active materials\,\cite{bastola2023formulation}, improving the understanding of material-interface behavior in multi-material prints\,\cite{carey2017fully}, and investigating the inkjet capability to build 3D-structures\,\cite{an2015high}. Quantum materials, including quantum dots and perovskite nanoparticles, have been printed using inkjet methods; this technique has created high resolution quantum dot light emitting diode (QLED) devices\,\cite{yang2020high}, and a heterogeneous perovskite/graphene photodetector\,\cite{austin2023photosensitisation}.

%[ALSO NEED TO ADD PARAGRAPHS ON FDM AND DIW - should say something about all methods that appear in the figure.]

\section{Design methods}

A major motivation for developing new systems through additive manufacturing is that the achievable designs are not limited by (traditional) manufacturing methods and can focus on the functional requirements needed for the task---for example following numerical simulation of the device with respect to external stress.
In\,\cite{AMUHV}, this was done for a 3D-printed vacuum chamber, where the deformation under an external pressure load was analyzed via ``Finite Element Analysis" (FEA) based simulation using ``MSc Marc 2017.1.0". Based on simulation insights, designs for AM production can be adapted with a high degree of freedom, for example to meet mechanical requirements with minimal addition of material, or to provide additional support along the direction of stress using advanced latticing\,\cite{lattice1, Maskery2018}. %Additively manufactured chambers can be built such that a requirement following a theorectical calculation can be implemented; the use of an additively manufactured chamber was suggested to trap domain walls in the search for dark matter \cite{Clements2024}.\\
AM can also take complex simulations/predictions from theory and directly turn them into an experimentally testable part, matching requirements in areas such as heat dissipation, eddy current response etc. A notable example is the use of AM to create a carefully tailored mass density distribution in order to experimentally test a class of dark matter models, which include domain walls that can be influenced by the local mass density \cite{Clements2024}. %Proposals such as this demonstrate how AM designs can open new opportunities for QT methods, not just enabling technological applications but also helping to address questions of fundamental scientific interest. 

The high degree of design freedom offered by AM leads to an increasing role for computational design methodologies\,\cite{design_optimisation_review}. These can involve top-down procedural algorithms, for example to map small-scale textures or lattices over complex surfaces\,\cite{AMUHV,conformal_mapping}, as well as standard or machine-learning based strategies for the optimization of one or more build parameters subject to known constraints\,\cite{design_optimisation1}. The application of such methods within QT will enable specific experimental needs to be met with maximum efficiency.

\section{Structural elements}

%Structural elements are a simple but essential part of all hardware, holding components in place and protecting them from mechanical damage. They are essential for quantum technologies, e.g. to deliver stable and reliable optical systems or correctly position coils and magnetic shields with complex geometries. 

Structural elements are typically used for assembly of components and their protection from mechanical damage. They are essential for quantum technologies, e.g. to deliver stable and reliable optical systems or correctly position coils and magnetic shields with complex geometries. 

%Elements outside the vacuum have been printed in polymers and ceramic materials, via both FDM and SLA methods, while elements for in-vacuum use have been additively manufactured in a variety of metals and alloys via LPBF, as well as in a more limited selection of polymer and ceramic materials, usually via SLA.%titanium, AlSi10Mg, stainless steel, Tullomer and others as discussed below.\\ %Even where such components could be produced conventionally, the utility of AM for rapid prototyping and the production of small numbers of intricate components has already enabled advances in the deployment of QT for real-world applications; for example, the success of recent work on brain pattern imaging with atomic magnetometers has depended in part on an additively manufactured helmet for appropriate positioning of the magnetometers in the array \cite{SPMIC}. There is no doubt that this kind of advantage will continue to be felt in research and development across the QT sector, but perhaps of greater interest are areas where the deployment of AM provides a more direct, functional benefit.  

All structural components benefited from mass reduction opportunities \,\cite{light_AM1,light_AM2,AMlightweight1}, including several in QT applications\,\cite{birmingham_flange,AMUHV}, achieved through ``structural optimization": the process of shaping a component so as to meet mechanical specifications with a minimum amount of total material, usually with the aid of computational design tools\,\cite{struc_opt1,struc_opt2}. This process often results in complex, organic-looking forms and/or intricate, small-scale patterning that would be highly impractical with conventional manufacturing; see, for example, the AM vacuum chamber shown in Figure\, \ref{fig:vacuum}(a). Usually, structural optimization relies on ``latticing" - the replacement of a solid volume of material with numerous, interconnected struts whose exact shape and density can be varied to meet specific requirements\,\cite{lattice1, Maskery2018}. %There have been several demonstrations of low-mass QT components using AM, some of which exploit structural optimisation .

 The role of structural components was expanded to, facilitate rapid heat dissipation from key areas\,\cite{heatdiss1,heatdiss2} and damping or isolating mechanical vibrations to improve stability\,\cite{vibdamp_thesis,vibdamp1,vibdamp2}. These capabilities have proved transformative for other sectors, and their wider deployment within QT is likely to have substantial impact on the sector. Below, we consider two areas of particular relevance to QT: optomechanics and vacuum equipment. 

%In addition to these general advantages, certain kinds of component have their own unique requirements that AM is particularly suited to meeting. Below, we consider two areas of particular relevance to QT: optomechanics and vacuum equipment. 

\subsection{Optomechanics and component positioning}

\begin{figure*}%[h!]
    \centering
    \includegraphics[width=1\textwidth]{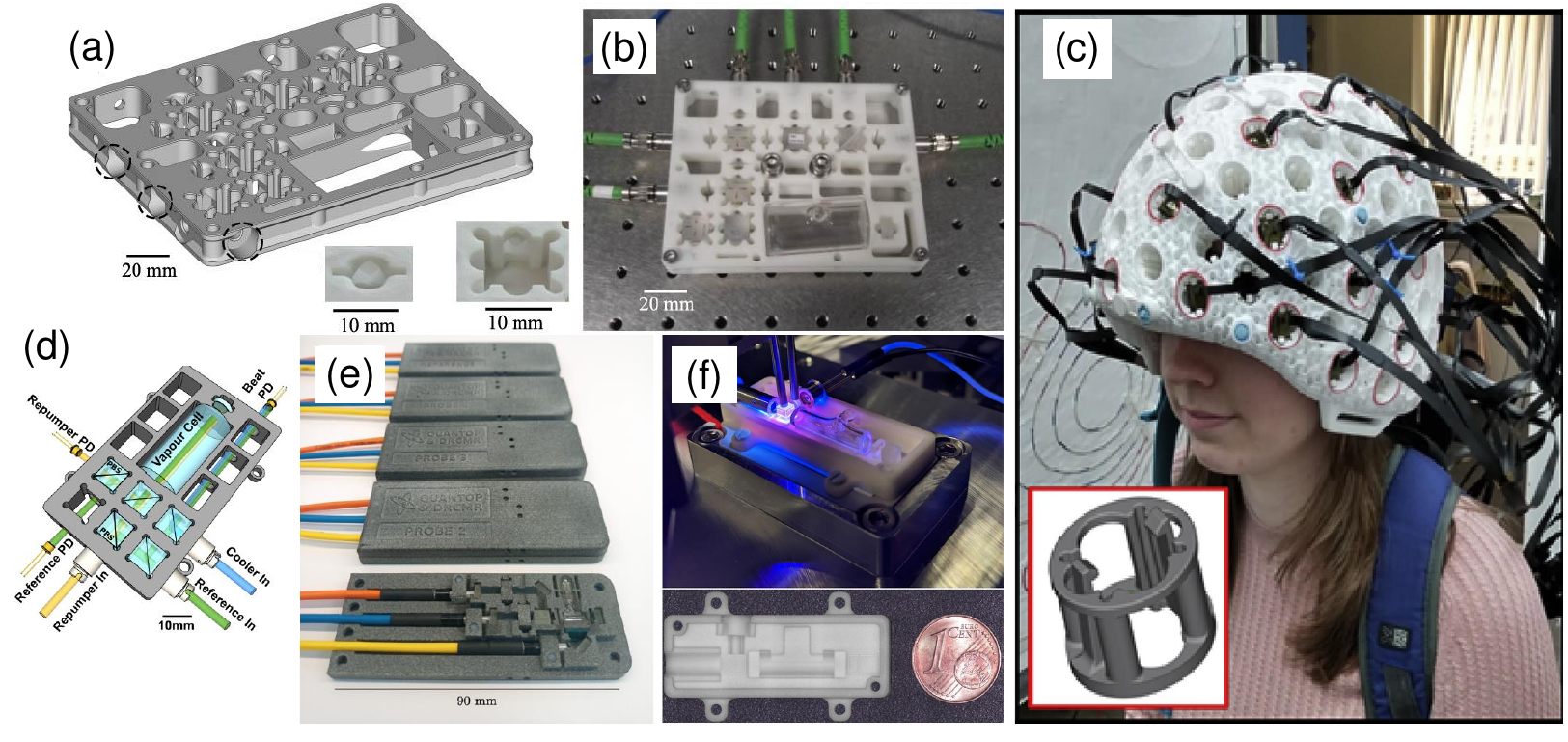}
    \caption{(a) Design of an AM optical mounting framework for compact magneto-optical trapping apparatus\,\cite{optamot}, including frequency stabilization for three wavelengths and power distribution into three optical fibers. (b) The final system printed in reinforced resin\,\cite{optamot} contains no movable elements apart from rotatable waveplates that are glued in place after alignment. (c) A 3D-printed helmet for exact positioning of optically-pumped magnetometer (OPM) devices\,\cite{Hill_2020}. (d) Compact spectroscopy and frequency stabilization system for three frequencies printed in polylactic acid via FDM\,\cite{optamot2}. (e) Magnetometer array for high precision spectroscopy\,\cite{Staerkind_2023} following the approach of\,\cite{optamot}. (f) Miniaturized spectroscopic reference apparatus that exploits a ceramic AM mounting structure, as described in\,\cite{ceramref}.}
    \label{fig:structural}
\end{figure*}

Many QT hardware components require accurate positioning, often within complex, multi-component systems. AM allows rapid and cost-effective production of intricate, bespoke mounting systems that accelerate prototyping and research. For example, the success of recent work on brain pattern imaging with atomic magnetometers has depended in part on a lightweight AM helmet\,\cite{SPMIC,Hill_2020} for appropriate positioning of the magnetometers on a mobile subject (see Figure\,\ref{fig:structural}(c)). The construction allows a highly robust and wearable OPM system with a total weight (including sensors) of 1,7\,kg. The rigid helmet was additively manufactured from PA12 (a nylon polymer) using an EOS P100 Formiga Laser Sintering machine. AM enables 133 cylindrical sensor mounts, eliminating motion of any sensor relative to all other sensors, while ensuring air circulation through a stable matrix-based gyroid triply periodic minimal surface lattice between the sensors. The relative locations and orientations of the sensor mounts are known to a high level of accuracy, because a complete digital representation of the helmet exists, and the tolerance of the 3D printing process is approximately 300\,$\mu$m. Another example is the magnetometer array described in\,\cite{Staerkind_2023} and shown in Figure\, \ref{fig:structural}(e), in which an AM mount ensures correct relative positioning of all system components. %\\
%The magnetometer arrays shown in \ref{fig:helmet}b) are used for high-field precision saturated absorption spectroscopy. Here, a 3D-printed nylon enclosure measuring $90\times33\times10$\,mm$^3$ is used produced via HP Multi Jet Fusion (MJF) technology. This material is compatible with temperatures approaching 100\,$^\circ$C during the measurements the probes have temperatures around 43$\,^\circ$C. 

%However, mass reduction through latticing is not the only motivation for AM of structural components. Another advantage of AM is its suitability for rapid production of small numbers of intricate, custom components. This can accelerate prototyping and enable very large SWAP reductions by allowing the use of bespoke structural components that fit the exact requirements of individual devices, something often highly impractical (if technically possible) with conventional manufacturing methods. 

Of particular relevance to QT is the positioning of optical components, such as mirrors and lenses, within the optical subsystems of QT devices. Here, not only the positions but also the \emph{orientations} of all components must be set and maintained with high precision---orientation control in such optical systems is often needed at the sub-milliradian level. AM shows particular promise in this area because the miniaturization enabled by AM reduces the stringency of orientational alignment requirements; compact AM versions of optical systems have been found to demonstrate superior stability in consequence\,\cite{optamot, optamot2, ceramref}.

The following formula for the approximate position deviation of a laser beam after passing through an optical system is given in\,\cite{optamot}:
\begin{equation}
\Delta r = \left[\Sigma_i (2\Delta \phi L_i)^2 \right]^{1/2},
\label{cumulative_deviation}
\end{equation}
where $\Delta r$ is the position uncertainty of the beam's center in some plane of interest, each reflective optic in the beam path is assumed to have independent alignment uncertainty $\Delta \phi$ and $L_i$ is the path length from the optic $i$ to the plane of interest. Where the scale of $\Delta \phi$ is independent of the size of the system this indicates a linear scaling of beam position uncertainty with optical path length; in reality the relative alignment deviations of components are likely to be greater for larger systems, due to the lower degree of curvature required in a mounting structure to achieve the same angular deviation when the spatial separation is greater. This formula thus illustrates the extreme importance of miniaturization in creating stable and robust optical apparatus, underpinning the success not only of\,\cite{optamot} but also of subsequent developments such as\,\cite{ceramref}, and highlighting the importance of miniaturization for achieving \emph{stable} QT devices as well as portable ones.

With this in mind, drastic miniaturization of such optical systems has been accomplished by embedding the optics within custom-built polymer\,\cite{optamot,optamot2} or ceramic frames\,\cite{ceramref}, where the suitability of AM for economic production of small batches of intricate components has facilitated both miniaturization and stability by eliminating the need for adjustable components in these mounting systems. In\,\cite{optamot2}, an atomic spectroscopy apparatus with a total volume of $<0.1$\,l and a mass of 120 grams is demonstrated (see Figure\,\ref{fig:structural}(d)), with a point of particular interest being that it is produced from polylactic acid (PLA) via FDM. Based on AM design and fabrication this device achieves remarkable stability. PLA and FDM are a common and economic AM material and method, and the relevant material properties of PLA, such as elastic modulus and coefficient of thermal expansion, are not particularly favorable for device stability. Despite this, the device demonstrated in\,\cite{optamot2} achieves far greater stability and robustness against environmental influence than standard lab systems; it is able to keep a laser frequency-stabilized to an atomic transition despite being subjected to temperature variations between 7 and 35 $^{\circ}$C and significant vibrational disturbances in a frequency range from 0 to 2 kHz. This illustrates how miniaturization and careful design, exploiting the freedom provided by AM, can overcome the limitations of build materials and deliver high overall performance with both low SWAP and low cost. 

Subsequent work has applied the same approach with more advanced materials and methods, leading to further performance enhancements. For example, in\,\cite{ceramref} a miniaturized spectroscopic frequency reference device is demonstrated with a total volume of 5.6\,ml and a mass of 15\,grams (Figure\,\ref{fig:structural}(f)). Laser frequency stabilization is implemented using the spectroscopic signal from this device and enables frequency instabilities as low as 1.5\,kHz over 100\,s, and $<40$\,kHz over 10$^4$\,s; stability is in part aided by the very short total optical path length of 53\,mm. The materials and methods demonstrated in \cite{ceramref} were subsequently adapted to demonstrate a highly-compact optical dipole trap for $^{87}$Rb atoms, a critical subsystem of many quantum sensing technologies \cite{Christ2024}.

\subsection{Vacuum components}

While vacuum apparatus is in essence a structural element, its use in high-vacuum applications places more stringent requirements on the methods and materials involved; vacuum apparatus must be leak-tight, able to make sealed connections to other vacuum components, have sufficient mechanical strength to survive robust assembly and atmospheric pressure, and be free from both trapped gas pockets and materials that outgas excessively. 

\begin{figure*}%[ht]
    \centering
    \includegraphics[width=\textwidth]{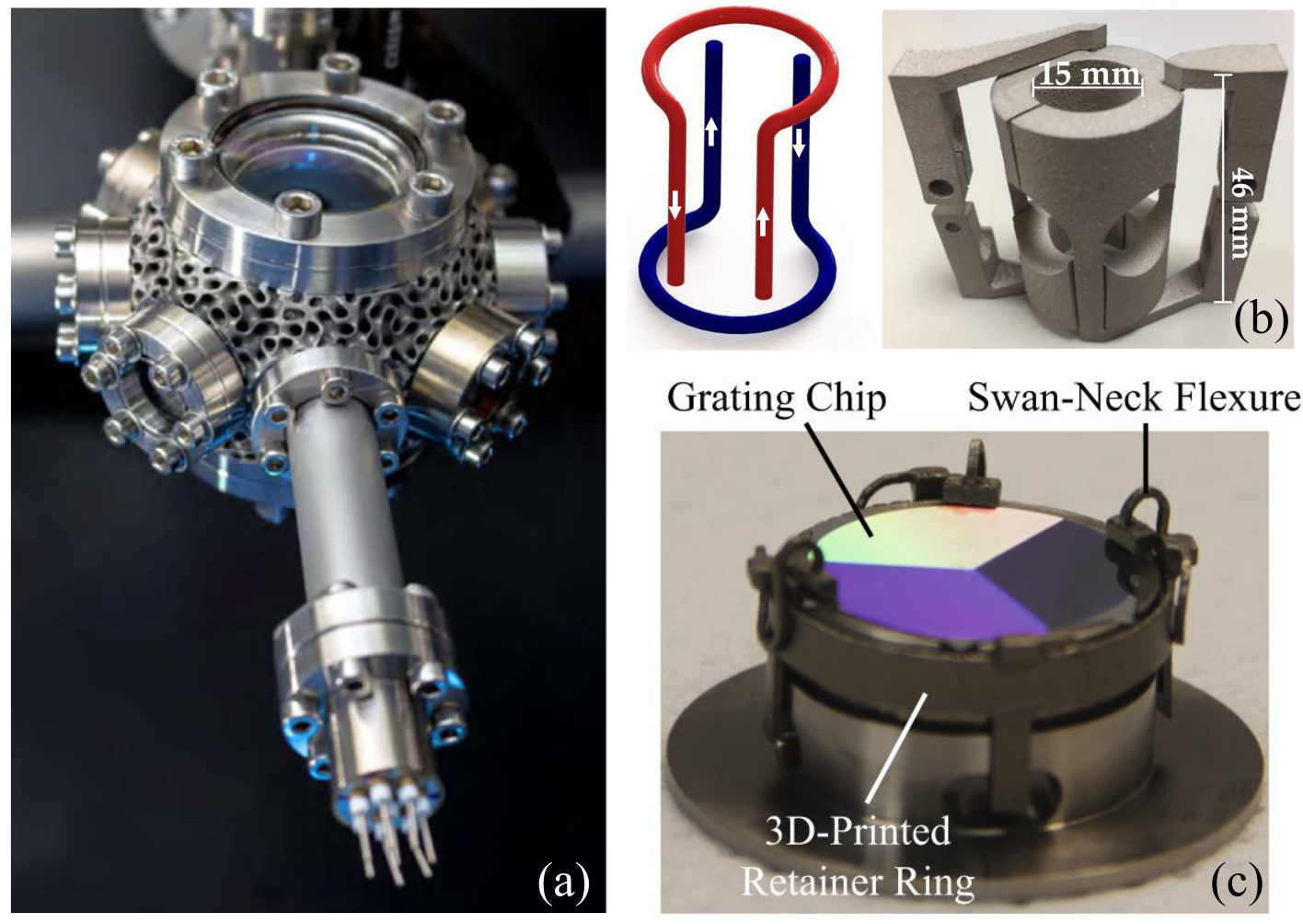}
    \caption{Additively manufactured components for UHV systems. (a) AM UHV chamber demonstrated in\,\cite{AMUHV}. Structural optimization, taking the form of a gyroid lattice, is visible on the chamber exterior and enables considerable mass reduction. (b) AM coils for a magneto-optical trap described in\,\cite{saint_coils} and sketch illustrating current flow during operation. (c) An AM grating mount for a single-beam magneto-optical trap\,\cite{swanneckGMOT}; AM enables a bespoke swan-neck flexure mount and avoids the use of epoxies.}
    \label{fig:vacuum}
\end{figure*}

The high and ultra-high vacuum (UHV) regimes are most relevant to QT, as maintaining quantum coherence requires minimal interactions with unwanted gas particles. AM was not initially considered appropriate for UHV due to its characteristic rough surfaces and porosity, and early attempts to apply AM to high-vacuum apparatus achieved only modest performance\,\cite{Jenzer_2017}. The first widely-publicized examples of AM components operating in the UHV regime occurred in 2018 with an AM-flange\,\cite{birmingham_flange} and 3D-printed in-vacuum coils\,\cite{saint_coils}, as seen in Figure\,\ref{fig:vacuum}(b). These were subsequently followed by the demonstration of a full vacuum chamber produced via powder-bed fusion\,\cite{AMUHV}. These demonstrations showed that porosity could be reduced to acceptable levels through appropriate build procedures and post-processing heat treatments, while surface roughness appeared not to be as destructive to vacuum performance as previously feared. 

% \begin{figure}[h!]
%     \centering
%     \includegraphics[width=0.5\textwidth]{Figure_printed_coils.pdf}
%     \caption{Additively manufactured coils for a magneto-optical trap described in \cite{saint_coils}.}
%     \label{fig:printedcoils}
% \end{figure}

% \begin{figure}[h!]
%     \centering
%     \includegraphics[width=0.5\textwidth]{urchin.png}
%     \caption{Additively manufactured UHV chamber demonstrated in \cite{AMUHV}. Structural optimisation, taking the form of a gyroid lattice, is visible on the chamber exterior and enables considerable mass reduction.}
%     \label{fig:urchin}
% \end{figure}

A number of different materials and methods have been explored for AM vacuum components. While\,\cite{saint_coils,AMUHV} used AlSi10Mg produced via laser powder-bed fusion, 
%from a powder-alloy using Al 88.9 wt\%, Si 10.7 wt\%, Mg 0.5 wt\%
high-vacuum components have also been additively manufactured in Titanium\,\cite{birmingham_flange,AMdispenser} and stainless steel\,\cite{Jenzer_2017,Jenzer_2019} via the same method and in glass\,\cite{wang2024additive} and aluminium oxide\,\cite{ceramref} using vat polymerization. Meanwhile, Ratkus et al. have obtained promising initial results for vacuum compatibility from test samples of AM copper, produced using LPBF \cite{Ratkus_2024}. 
Though techniques involving a binding agent, such as binder jet printing\,\cite{binderjet}, have not yet been as successfully applied to metal AM for UHV, the above demonstrations of ceramic components fabricated using ceramic-doped polymers indicate that the use of a removable binding agent does not preclude a method for UHV applications. 

Recently, the suitability of printed plastics for HV and UHV has been explored, with a number of materials fabricated via FDM and mSLA found to be suitable for deployment in HV and even UHV conditions\,\cite{tullomer,photopolymer_AM}. In particular, the proprietary material Tullomer was found to exhibit desorption rates as low as 5$\times10^{-11}$\,mbar\,l\,cm$^{-2}$s$^{-1}$\,\cite{tullomer}, with vessels containing printed test-samples reaching pressures below 10$^{-9}$\,mbar. 

The application of thin, UHV-compatible coatings to more traditional AM materials has also been found to be effective in improving vacuum performance\,\cite{vac_coating}. Evidence was found in\,\cite{AMUHV} to suggest that the material outgassing rate was being reduced by a protective surface layer, indicating that this approach is applicable across different materials and AM methods. %Indeed, in \cite{AMUHV}, x-ray photoelectron spectroscopy was used to assess the chemical composition of the surface layers of the as-printed material. Although the mechanism by which it occurs is not yet known, clear evidence was found of Magnesium enrichment in the surface of the printed material. Mass spectrometry measurements then revealed that heating to $>300^{\circ}$C under vacuum irreversibly increased the outgassing rate of the material, lending weight to the hypothesis put forward in \cite{AMUHV} that a protective surface layer of MgO forms on the material and inhibits outgassing processes. This is thought to explain the unexpectedly good vacuum performance of the material observed in \cite{AMUHV}, and is further evidence that protective surface coatings can enable a wider range of AM materials to be deployed in UHV applications. 

The potential advantages of AM for vacuum components are numerous, and perhaps the best way to review them is to examine representative examples of each case. There is, naturally, significant overlap with the benefits of AM for other structural components. Light-weighting has been demonstrated in\,\cite{AMUHV}, where a structure feasible only via AM enabled a mass reduction of approximately one third (the chamber in fact weighed around 70\% less than industry-standard equivalents, but not all of this advantage stemmed from AM-specific developments). Meanwhile, \cite{saint_coils} illustrated the benefits of rapid prototyping and the ease of fabricating custom components to meet specific requirements, using AM to produce a set of coils for magneto-optical trapping with unusually low power consumption and switching time. While conventional fabrication of this coil structure would have been possible, AM enabled rapid production and testing of an unusual structure specific to the needs of an individual experiment. 
The AM AlSi10Mg that comprised these coils was produced using laser powder bed fusion and yielded a cold resistivity of $5\times 10^{-8}~\Omega$m, or a conductivity that is 70\% of the bulk material. 

In\,\cite{swanneckGMOT} a grating chip used for magneto-optical atom trapping was held in place with an additively manufactured retainer that has four swan-neck flexure mounts using Ti-alloy (ASTM Ti-6Al-4V)---see Figure\,\ref{fig:vacuum}(c). The flexure mounts provided excellent optical access to the surface of the grating chip and avoided the use of epoxies.

% \begin{figure}[h!]
%     \centering
%     \includegraphics[width=0.5\textwidth]{Figure_printed_grating_mount_v1.pdf}
%     \caption{An additively manufactured grating mount for a single-beam magneto-optical trap \cite{swanneckGMOT}; AM enables a bespoke swan-neck flexure mount and avoids the use of epoxies.}
%     \label{fig:printedcoils}
% \end{figure}

Beyond these demonstrated examples, additive manufacturing of UHV components promises a range of further benefits for QT. One is part consolidation. Most current UHV systems are assembled from modular components to enable the construction of bespoke systems from industry standard parts that can be efficiently manufactured using conventional techniques. Because of the need to maintain a vacuum seal, the connections between these modular components are bulky and laborious to assemble; eliminating them through part consolidation facilitates rapid prototyping and leads to considerable miniaturization and mass reduction. Consolidation also aids reliability by reducing the number of potential leak locations. AM allows efficient production of customized systems, removing the need for modular components and thus enabling part consolidation. Other potential areas of benefit include selectively screening key experimental regions from known outgassing sources, finely controlling gas flow and channel conductance and tailoring the eddy current response of conductive structures to changes in magnetic field. Ongoing work by the authors is also showing that AM-enabled surface patterning can improve the effectiveness of passive pumping systems based on non-evaporable getters \cite{surfacepaper}. All of these have clear usage cases within QT and stand to be enabled or facilitated by the advancement of AM for UHV applications.

\section{Transparent elements and optics}

\begin{figure*}%[h!]
    \centering
    \includegraphics[width=\textwidth]{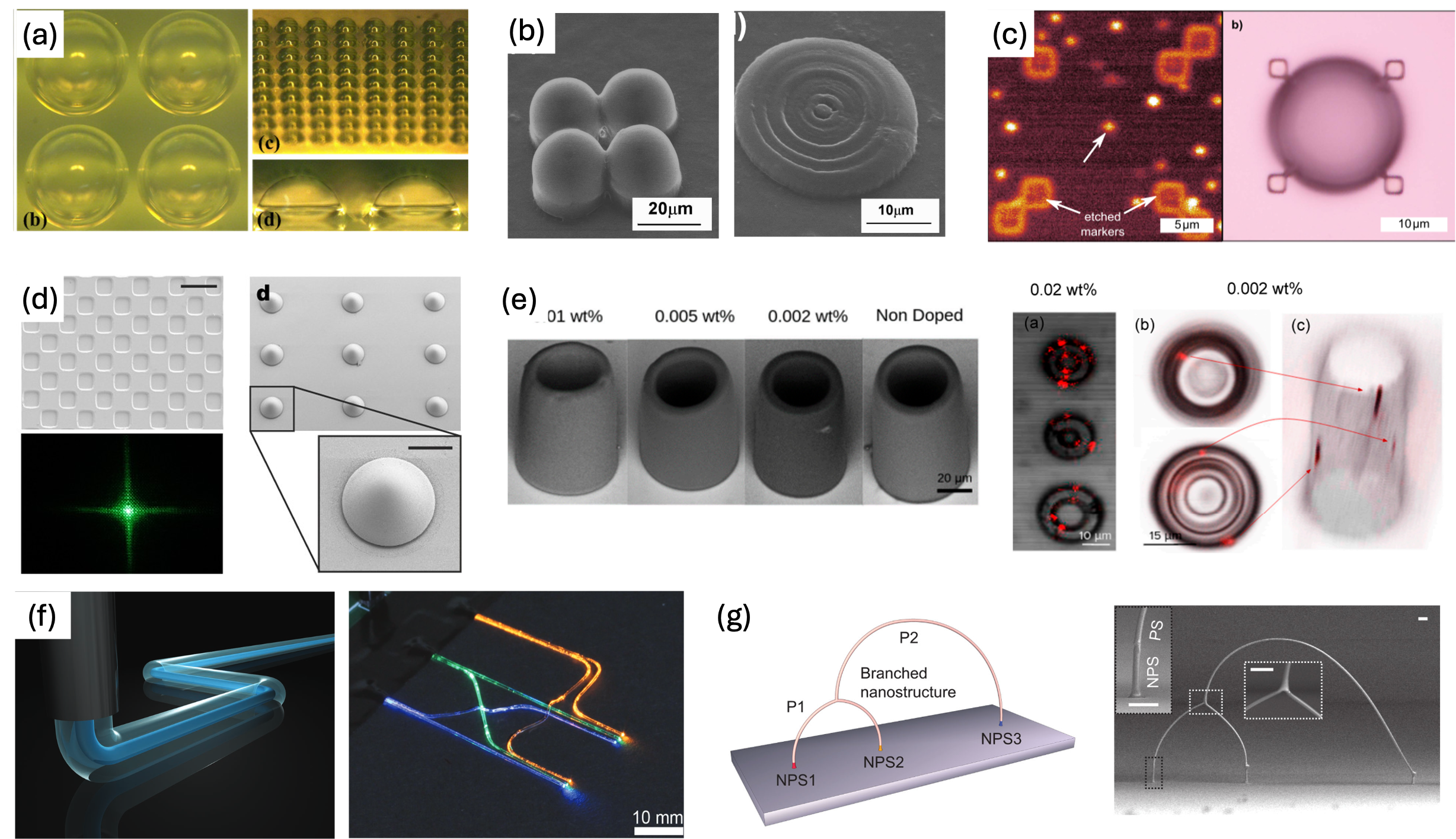}
    \caption{Examples of transparent optical components by AM for quantum technologies.
    (a) Micro-lens array of average diameter 333.28 $\mu$m by inkjet printing \cite{xie2012fabrication}.
    (b) Spherical micro-lenses (left) and micro Fresnel lens fabricated by two-photon polymerization \cite{guo2006micro}.
    (c) Photoluminescence intensity map (left) and the optical microscope image (right) of solid immersion lens printed on the single quantum dot for light coupling \cite{sartison2017combining}.
    (d) Glass micro-optical diffractive structure with the optical projection pattern (left) and microlenses fabricated using micro-stereolithography. (scale bar: $100\,\mu$m).  Reprinted with permission from\,\cite{kotz2017three}.
    (e) Left: scanning electron microscope image of cylindrical microstructues fabricated using two-photon polymerization, doped with different proportions of nanodiamonds. Right: confocal scanning image and 3D rendering of structures doped with 0.02 wt\% nanodiamond showing fluorescent spots artificially coloured \cite{couto2023integrating}.
    (f) Left: schematic image of a direct-write assembly of photocurable liquid core-fugitive printed polymeric optical waveguides. Right: an optical waveguide network composed of six waveguides coupled with three LEDs that distributed colored light with minimal crosstalk. Reprinted with permission from\, \cite{lorang2011photocurable}.
    (g) Schematic diagram and scanning electron microscopy images illustrating a freestanding branched structure integrated on three different nanophotonic sources by meniscus-guided 3D printing. The white scale bars in the right-hand panel are 2 $\mu$m in length.  Reprinted with permission from\,\cite{pyo20163d}.
    }
    \label{fig:lens}
\end{figure*}

Transparent optical and photonic components are essential in quantum technology applications for beam shaping, collimation, focusing, transmission and imaging. With its expanding material libraries encompassing polymers, silica, and low dimensional materials, AM offers a flexible approach for producing complex micro-optics. These include lenses, waveguides and coupling interfaces for integration, with applications in quantum sensing, information processing and more. For instance, micro-lens arrays of different sizes and geometries have been fabricated using both inkjet printing and SLA.
Inkjet printing shows great potential for fabricating polymer convex or hemispherical refractive micro-lens arrays with diameters from 100 $\mu$m to several mm, offering high flexibility, minimal processing complexity and low production cost. Xie et al. have fabricated an array of $9 \times 9$ polymeric micro-lenses with a diameter of 333.3\,$\mu$m, a pitch of 354.04\,$\mu$m, a sag height of 94.13\,$\mu$m and surface roughness of 0.243\,nm using UV curable polymers\,\cite{xie2012fabrication}. 
The pitch and the geometry of the micro-lens array are uniform (Figure\,\ref{fig:lens}(a)). 
For micro-lenses of $< 50\,\mu$m diameter, digital light processing methods are advantageous. A $2 \times 2$ micro-spherical lens array with diameter of 15\,$\mu$m and a micro Fresnel lens with diameter of 17\,$\mu$m have been been fabricated by Guo et al. using two-photon polymerization, as shown in Figure\,\ref{fig:lens}(b)\,\cite{guo2006micro}. Here a SCR500 acrylate resin was used with a transparency in the visible range of 85-90\%.

Similarly, internal reflection solid immersion lenses can be fabricated on top of single InAs quantum dots\,\cite{sartison2017combining,Sartison2021} and have increased the collection efficiency of the output photons by up to 9 times when aligned with gold markers (Figure\,\ref{fig:lens}(c)). The single-photon character of the source is preserved after device fabrication, reaching a degree of second-order coherence value of approximately 0.19 under pulsed optical excitation. The printed lens device can be further joined with an optical fiber and permanently fixed to the fiber\,\cite{Sartison2021}.

Limitations for transparent polymers and oligomers relating to chemical compatibility, vacuum compatibility, temperature tolerance, material leeching\,\cite{dragone20133d} and air-permeability pose challenges for QT applications. Glass is often a more desirable material due to its chemical inertness, high transparency across a wide spectral range and physical tolerances. AM of glass typically involves a fumed silica nano-powder resin which is then thermally processed after printing to remove unwanted compounds and sinter the aggregated powder into transparent glass. Kotz et al.\,\cite{kotz2017three} have demonstrated micro-optical diffractive elements and glass micro-lens arrays by micro-stereolithography, with a surface roughness of about 2\,nm (Figure\,\ref{fig:lens}(d)). The structure exhibits \>90\% transparency for wavelengths from UV to near-infrared. 

Functional materials for active QT-components can be integrated into micro-optical structures. For example, Nitrogen vacancy (NV) defects in diamond have been demonstrated as quantum sensors\,\cite{flinn2023nitrogen,radu2019dynamic}, where their properties are affected by environmental factors such as magnetic field, pressure or temperature. Nanodiamonds have been incorporated into additively manufactured components\,\cite{mangal2020incorporating}, but nanodiamond loading has to be well-controlled to prevent structural deformation due to light scattering during the printing process. For 0.002\,wt\% loading, NV-nanodiamond clusters can clearly be seen in AM structures fabricated via two-photon polymerization (Figure\,\ref{fig:lens}(e)\,\cite{couto2023integrating}.

Apart from micro-lenses, waveguides can also be fully fabricated by AM and might be used in the future for quantum sensing\,\cite{DaRos2020} or quantum computing\,\cite{Walmsley2013,Cooper2019}. Hybrid organic–inorganic optical waveguides have been demonstrated in straight, curved, and out-of-plane conﬁgurations via direct-write assembly of photo-curable liquid core-–viscoelastic fugitive shell inks by Lorang et al., as shown in Figure\,\ref{fig:lens}(f)\,\cite{lorang2011photocurable}. The printed waveguides exhibit nearly cylindrical morphology and low optical loss throughout the visible spectrum. For nanoscale photonic interconnections, Pyo et al. reported a meniscus-guided 3D printing method that allows direct, non-destructive integration of nanowire waveguides, as shown in Figure\,\ref{fig:lens}(g)\,\cite{pyo20163d}. No substrate leakage and excellent stretchable operations were observed for functional photonic components of multiplexers and splitters. 

\begin{figure*}%[h!]
    \centering
    \includegraphics[width=\textwidth]{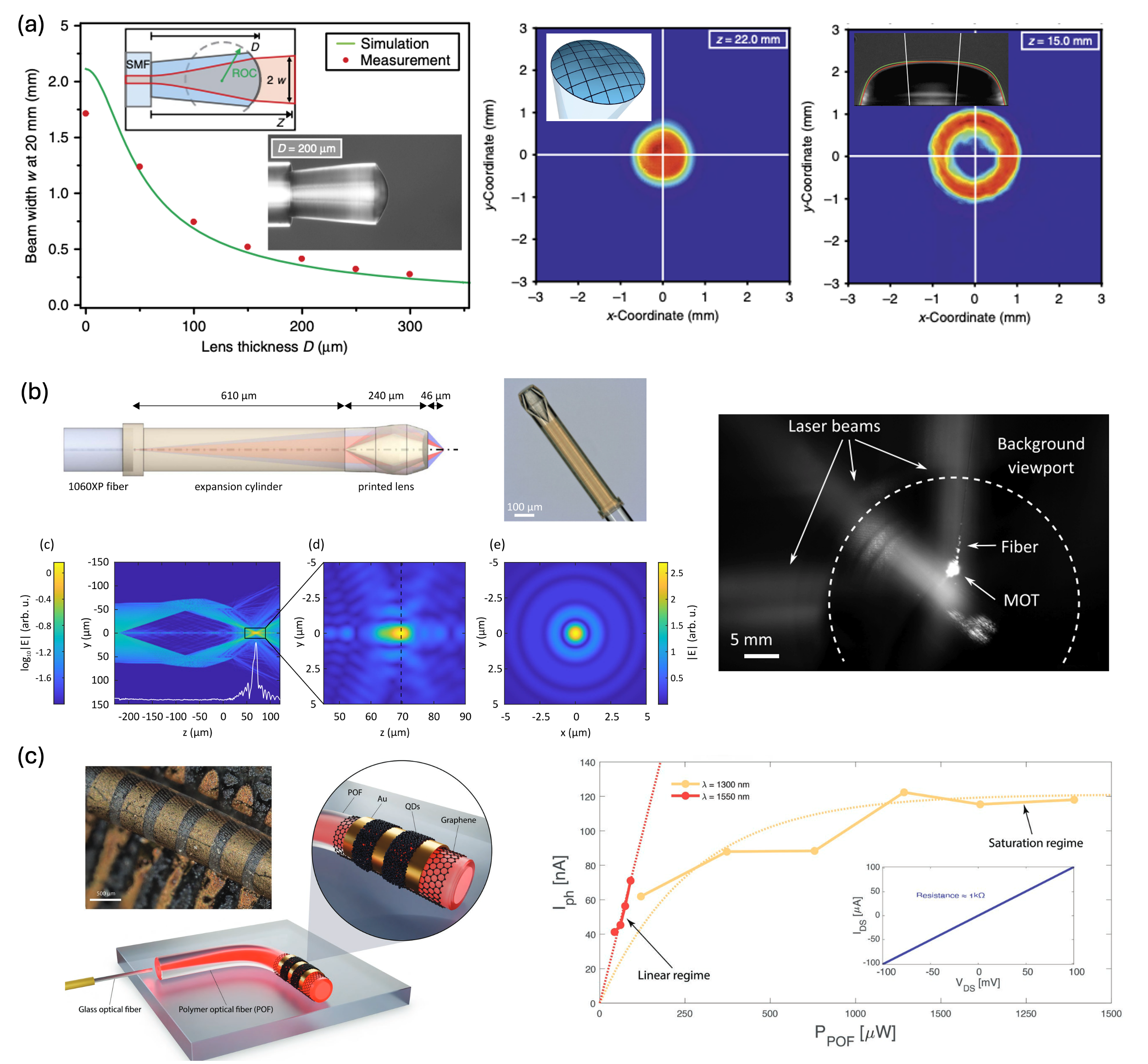}
    \caption{Printing components on optical fibers for quantum technology applications. (a) Left: design and optical microscope images of spherical lenses fabricated by direct laser writing with thicknesses ranging from 0 to 300\,mm and the corresponding beam width. Center: donut shaping free-form lens and the measured light distribution. Right: top-hat light shaping free-form lens and the measured light distribution \,\cite{gissibl2016sub}.
    (b) Left: a total internal reflection lens printed at the tip of a fiber and the wave propagation method simulated electric field with a zoom into the focal region and the radial cross-section at the focal spot. Right: creation of a MOT 125\,$\mu$m from the fiber tip with the TIR lens when positioned in the center of a UHV chamber \cite{ruchka2022microscopic}.
    (c) Left: schematic and optical microscope image of the hybrid-PbS quantum dot phototransistor printed on polymer optical fiber. Right: the photocurrent of the phototransistor under different light power for 1300 and 1550\,nm wavelengths\cite{kara2023conformal}.
    }
    \label{fig:fibres}
\end{figure*}

Micro-lenses have also been directly fabricated on the end facet of single-mode optical fiber to collimate a Gaussian beam emerging from the fiber. Spherical lenses with different radii of curvatures and thicknesses have been produced by Gissibl et al.\,\cite{gissibl2016sub}, the schematic shape and the resulting beam waist has been studied, showing that the collimation improves with increasing lens thickness (Figure\,\ref{fig:fibres}(a)). Additionally, the measured intensity distribution was used to infer the real surface profile and thus, via an iterative optimization algorithm, to improve the fabrication process. Furthermore, micro-lenses printed onto the tip of standard optical fibers have enabled a new trapping concept for ultracold atoms in optical tweezers\,\cite{ruchka2022microscopic}. The unique properties of these lenses make them suitable for both trapping of small numbers of atoms and capturing their fluorescence with high efficiency. In an exploratory experiment by Ruchka et al. (Figure\,\ref{fig:fibres}(b))\,\cite{ruchka2022microscopic}, the vacuum compatibility and robustness of these structures were demonstrated, and a magneto-optical trap for ultracold atoms was successfully formed approximately 125 $\mu$m from the fibre tip.

\section{Active components} 

\begin{figure*}%[h!]
    \centering
    \includegraphics[width=\textwidth]{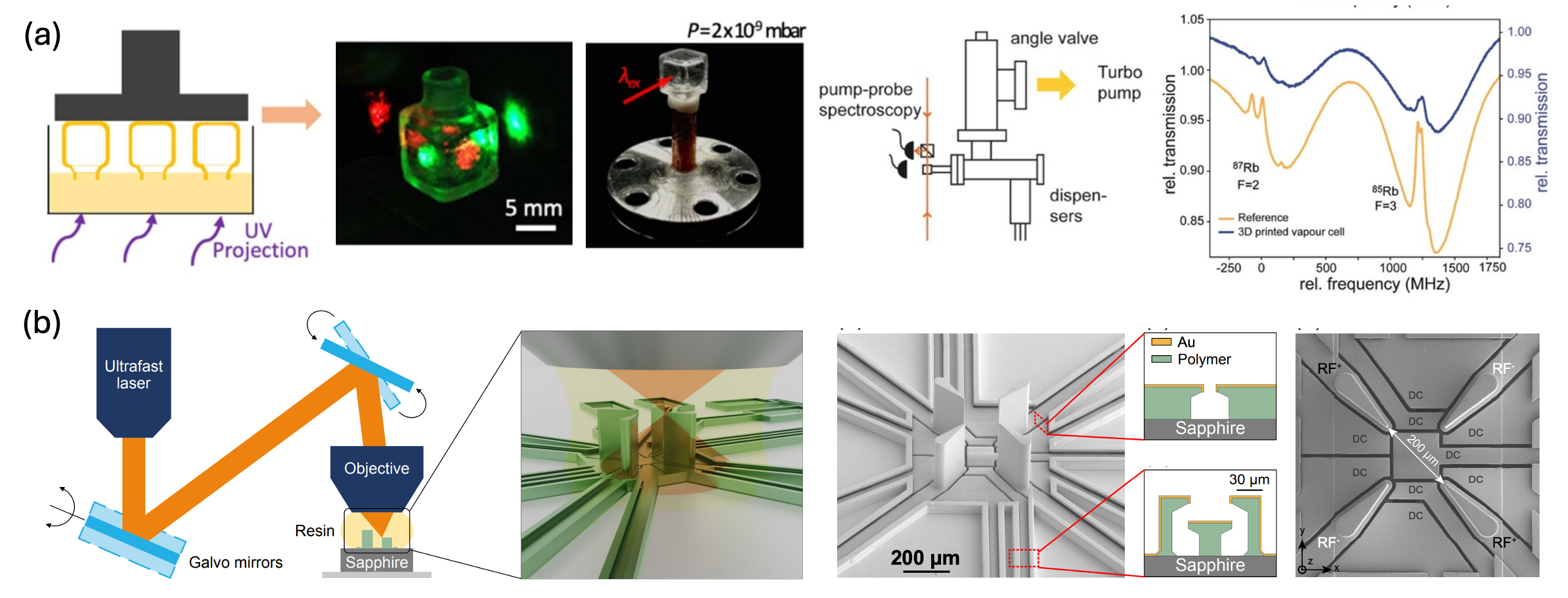}
    \caption{Examples of active components made by AM for quantum technology applications.
    (a) An glass vapor cell fabricated using vat polymerization, with cuboid shape and two clear optical axes. The cell is mounted to a standard UHV system to demonstrate Rb hyperfine spectroscopy\,\cite{wang2024additive}.
    (b) Illustration and scanning electron microscope images of a 3D-printed  vertical linear Paul trap, with coated gold electrodes for RF and DC voltages to achieve ion-trap quantum computing\,\cite{xu20233d}.    
    }
    \label{fig:active}
\end{figure*}

%functionalisation of optical fibres
The scope of AM for quantum technologies can be greatly expanded by incorporating functional materials for active components into the printed structure. Recently, an atomic vapor cell that was printed using vat polymerization\,\cite{wang2024additive} has been demonstrated. This cell, measuring 7\,mm per side, was pumped to $2 \times 10^{-9}$ mbar, maintaining this pressure with no noticeable leaks, before being filled with Rb vapor (Figure\,\ref{fig:active}(a)). A laser was passed through the cell and the resulting spectroscopy was used to frequency stabilize a laser to the $F = 2 \rightarrow F' = 3$ transition of $^{85}$Rb. The Allan deviation of the error signal was measured for a range of time scales, yielding a frequency instability of $\Delta F/F = 2 \times 10^{-10}$ for integration times of 1\,s. This performance was found to be approximately the same as the frequency stability achieved with a laser locked to a Thorlabs borosilicate cell, when normalized for the difference in optical path length. Integrated multi-material components, such as inkjet printed graphene and silver electrodes, are also demonstrated in\,\cite{wang2024additive}; these and similar elements are promising for applications such as photon detection, localized heating and active magnetic shielding. %There is further demonstration to integrate multi-material components, such as inkjet-printed graphene and silver electrodes for heating, photon detection, and potentially magnetic field shielding. 

Another key active QT component recently demonstrated is a 3D-printed micro ion trap for quantum information processing by Xu et al.\,\cite{xu20233d}. The trap is fabricated by first creating 3D polymer structures using two-photon polymerization and then coating the polymer with a 1\,µm-thick gold layer using electron-beam evaporation (Figure\,\ref{fig:active}(b)). The trap consists of four RF electrode pillars on a sapphire substrate with a total height of $300\,\mu$m. Nine planar DC electrodes are placed on the substrate surface for tuning the potential in a $600~\mu\mathrm{m} \times 600~\mu \mathrm{m}$ square. Single calcium ions have been confined in the 3D printed trap with radial frequencies ranging from 2\,MHz to\,24 MHz under room temperature, enabling high-fidelity coherent operations on a Ca$^{+}$ optical qubit after only Doppler cooling. The design freedom of AM offers an opportunity to scale this initial demonstration into multi-dimensional ion trap arrays with optimized performance and expanded functionality. Underpinned by integrated photonic circuits based upon AM waveguides, this architecture may have significant potential for scalable quantum computing. %The design freedom of AM is drastically expanded with scalability and precision for large scale device with integrated photonic circuits so that ion trap geometries can be optimised for higher performance and better functionality.

%\section{QT or other advanced methods for AM}

\section{Conclusion and Outlook}

AM has a demonstrated ability to substantially advance quantum technologies by enabling the fabrication of intricate, highly customized and scalable components. This review has highlighted the diverse applications of AM in quantum technology devices demonstrated to date, including structural components for assembly and high-vacuum apparatus, optical and photonic components, and functionalized active components. Quantum technologies can benefit in multiple ways from AM, including via reductions in SWAP, improved stability, and the enabling of integrated, multi-material structures not possible through conventional manufacturing, such as those demonstrated in \cite{wang2024additive,xu20233d}. Integration of the advantages of AM with the needs of QTs could bring transformative changes in sectors from healthcare and navigation to computing and communication. %IntegraThe ability to integrate multi-material architectures such as conductive traces, micro-optics and photonic waveguides demonstrates AM’s transformative role in manufacturing for quantum sensing, computing, and communication.%for applications demanding UHV compatibility and low-loss optical performance. The development of novel printable materials, and multi-material deposition strategies, will be essential for expanding AM’s applicability in quantum devices.
Particularly important are developments in the quantum sensing area, where AM of components for high vacuum applications was demonstrated with metal, glass and even polymers to produce parts with a high degree of design freedom, tailored to specific applications \cite{AMUHV,AM_VC}. \\
Quantum sensing is a fast growing area, which is currently largely limited to high value applications due to complexities in device manufacturing, operational environmental requirements and the advanced operator skill set needed; a manufacturing approach that focuses on design freedom and the efficient realization of complex parts, rather than on large production volumes, is urgently needed to enable these transformative applications. AM will be of increasing importance to the deployment of quantum sensors in the near future. % with multi-material AM showing particular promise for efficient fabrication of more complex, multi-component systems and devices.%hence development of manufacturing approached is essential for harnessing the ‘quantum advantage’ for transformative applications. 
Current research addresses open questions on the long-term stability of components, the fabrication precision of AM for certain techniques and material compatibility  \cite{derby2010inkjet, jensen2023long}. 
In this fast-evolving field we are witnessing the development of novel printable materials, deposition strategies and multi-material AM \cite{bastola2023formulation}. Furthermore, hybrid approaches that combine AM with traditional microfabrication techniques can enhance the performance and reliability of quantum components \cite{valentine2017hybrid,krimpenis2022application}.
%Looking ahead, the application of AM for emerging quantum technologies presents an exciting opportunity. The integration of 3D-printed components with quantum optics, atomic systems, and other light-matter interaction platforms has the potential to revolutionise the miniaturization and scalability of quantum hardware. Furthermore, the ability to fabricate quantum devices on demand using AM could accelerate the commercialisation of quantum technologies, improving device performance, reducing costs and facilitating scalable production. Continued interdisciplinary collaboration will be critical in overcoming existing technical barriers and realising the full potential of AM in quantum technologies.
Looking ahead, AM has the potential to provide an ideal manufacturing solution for emerging quantum technologies and enable the widespread use of quantum systems outside the laboratory, notably in key environments such as in space, on vehicles or underground. Combining the benefits of optimized design with on-site, on-demand manufacturing of compact and packaged field-deployable devices, AM will help to accelerate the development and commercialization of quantum technologies. However, the development of AM supported quantum technologies has only just begun, and key foundational advancement might await in the near future, in the development of the interface between quantum science and technology and the materials and processes of additive manufacturing: for example via research into AM of active quantum materials, complete QT-AM sensors or sensor arrays or AM supported quantum computers or optimization of AM materials and processes with respect to the sensitivity and long-term precision and stability of AM fabricated quantum sensors.
Continued interdisciplinary research is therefore critical in overcoming existing technical barriers and supporting the realization of the transformative promise held by quantum technologies.

\bigskip

% Acknowledgments
\noindent \textbf{\large Acknowledgments}
\smallskip

\noindent  
We acknowledge supported by the grant 62420 from the John Templeton Foundation, the IUK project No.133086, EPSRC grants EP/T001046/1, EP/R024111/1, EP/M013294/1, EP/Y005139/1 and  EP/Z533166/1.

%future impact

%combination of materials

%fibres to be integrated with QT devices
%electronics and optics by AM

%https://www.lboro.ac.uk/research/amrg/research/additive-manufacturing-for-quantum-systems/

%https://sd.llnl.gov/article/23291/additive-manufacturing-meets-quantum

%https://www.manufacturing-quality.com/quality-news/qant-quantum-sensor-additive-manufacturing-particle-analysis/

%3D-Printed Micro Ion Trap Technology for Scalable Quantum Information Processing
%https://arxiv.org/pdf/2310.00595  \\

%Additively Manufactured Ceramics for Compact Quantum Technologies
%https://onlinelibrary.wiley.com/doi/full/10.1002/qute.202400076  \\

%Complex Three-Dimensional Microscale Structures for Quantum
%Sensing Applications
%https://pubs.acs.org/doi/full/10.1021/acs.nanolett.3c02251 \\

%High-kinetic inductance additive manufactured superconducting microwave
%cavity
%https://pubs.aip.org/aip/apl/article/111/20/202602/34412 \\

%https://www.nano-di.com/resources/blog/the-role-of-3d-printed-electronics-in-revolutionizing-quantum-sensor-research-for-universities-and-research-labs

\bibliography{sample}

\end{document}